\documentclass[12pt]{article}
\renewcommand{\baselinestretch}{1.09}
\usepackage{graphicx}
\begin{document}
%
\begin{flushright}
  OU-HET-549 \ \\
\end{flushright}
\vspace{0mm}
\begin{center}
\large{GPDs and underlying spin structure of the nucleon}
\end{center}
\vspace{0mm}
\begin{center}
M.~Wakamatsu\footnote{Email \ : \ wakamatu@phys.sci.osaka-u.ac.jp}
\end{center}
\vspace{-5mm}\begin{center}
Department of Physics, Faculty of Science, \\
Osaka University, \\
Toyonaka, Osaka 560-0004, JAPAN
\end{center}

\vspace{-2mm}
\begin{center}
\small{{\bf Abstract}}
\end{center}
\vspace{0mm}
\begin{center}
\begin{minipage}{15.5cm}
\renewcommand{\baselinestretch}{1.0}
It is shown that, based only upon two empirically known facts
besides two reasonable theoretical postulates, we are inevitably led
to a model-independent conclusion that the quark orbital angular momentum
carries nearly half of the total nucleon spin at the low energy scale
of nonperturbative QCD. Also shown are explicit model predictions
for the forward limit of the unpolarized spin-flip GPDs, which are
believed to give valuable information on the distributions of quark
angular momentum inside the nucleon.
\end{minipage}
\end{center}


 \section{\large{Introduction}} 

The so-called nucleon spin puzzle raised more than 15 years ago
is still an unsolved fundamental puzzle in hadron physics \cite{EMC88}.
If intrinsic quark spin carries little of total nucleon spin,
what carries the rest of the it ? That is the question to be
answered. Admitting that QCD is a correct theory of strong
interactions, the answer must naturally be searched for in some
one of the following three, i.e. the quark orbital angular
momentum (OAM), the gluon polarization or the gluon orbital
angular momentum.

An important remark here is that it has little meaning to talk about
the spin contents of the nucleon without reference to the energy scale
of observation. In fact, it is a widely known fact that the gluon
polarization grows rapidly as $Q^2$ increases, even if it is small
at low energy. Conversely, the gluon orbital angular momentum
decreases rapidly to partially compensate the increase of $\Delta g$.
Hence, when we talk about the nucleon spin contents {\it naively},
we should understand that we are thinking of it at low energy scale of
nonperturbative QCD.

Roughly speaking, there exist two opposing or contrasting standpoints
to try to answer the above question. The chiral soliton picture of
the nucleon emphasizes the importance of the quark orbital angular
momentum \cite{BEK88},\cite{WY91}\nocite{WK99}--\cite{WW00}.
We recall that, the dominance of the quark OAM in these
unique models can be traced back the collective motion of quark
fields in the rotating hedgehog mean field \cite{WY91}.

On the other hand, the possible importance of gluon polarization
was emphasized by several authors on the basis of the axial
anomaly of QCD \cite{AR88}\nocite{CCM88}--\cite{ET88}.
Later, the role of QCD anomaly was understood more
precisely at least within the framework of perturbative QCD.
That is, the perturbative aspect of axial anomaly is understood as
a factorization scheme dependence of the longitudinally polarized
PDF in the flavor singlet channel. However, the nonperturbative
aspect of it is left totally unresolved. As a consequence, no one can
give any reliable theoretical prediction for the actual magnitude
of $\Delta g$. Probably, one of the most promising attempts aiming
at a direct measurement of $\Delta g$ is to use photon-gluon fusion
processes. For instance, the Compass group recently extracted the value of
$\Delta g / g$ from the analysis of the asymmetry of high $p_T$ hadron
pairs \cite{Compass05}.
Their first result for $\Delta g / g$ has turned out fairly small,
$\Delta g / g \ \sim \ 0.06 \ \pm \ 0.31$,
although it would be premature to draw any decisive conclusion only
from this result.

On the other hand, the key quantity for the direct
measurement of $J_q$ and/or $L_q$ is the generalized parton distributions
(GPDs) appearing in the cross sections of deeply virtual Compton
scattering and deeply virtual meson productions. As is widely known,
what plays the central role here is Ji's quark angular momentum
sum rule \cite{Ji98}.

 \section{\large{Generalized form factors and quark orbital angular
 momentum}}

Here, let us start with the familiar definition of generalized form
factors $A_{20}(t)$ and $B_{20}(t)$ of the nucleon, which is given as
a nonforward nucleon matrix element of QCD energy momentum tensor :
\begin{equation}
 \langle N(P') |\, T^{\mu \nu}_{q,g} \,| N(P) 
 \rangle \ = \ \bar{U} (P') \left[ \, 
 A_{20}^{q,g} (t) \,
 \gamma^{( \mu} P^{\nu )} \ + \ 
 B_{20}^{q,g} (t) \,
 \frac{P^{( \mu} i \sigma^{\nu ) \alpha} 
 \Delta_\alpha}{2 M} \right] \,U(P) .
\end{equation}
The famous Ji's sum rule relates the total angular momentum carried
by quarks and gluons to the forward limit of these generalized
form factors \cite{Ji98} :
\begin{eqnarray}
 J^{u + d} &=& \frac{1}{2} \,
 \left[ \,A_{20}^{u+d} (0)
 \ + \ 
 B_{20}^{u+d}(0) \,\right] ,\\
 J^g &=& \frac{1}{2} \,
 \left[ \,A_{20}^g (0)
 \ \ + \ \ 
 B_{20}^g (0) \,\right] .
\end{eqnarray}%
Here, the first $A_{20}(0)$ parts reduce to the total momentum
fractions of quarks and gluons as
\begin{eqnarray}
 A_{20}^{u+d}(0)
 &=& \int_0^1 \left[ \,
 u(x) + \bar{u}(x) + d(x) + \bar{d}(x) \,\right] \,dx
 \ \equiv \ {\langle x \rangle}^{u+d} , \\
 A_{20}^g (0) 
 &=& \int_0^1 x \,g(x) \,dx
 \equiv \ {\langle x \rangle}^g ,
\end{eqnarray}
while the second $B_{20}(0)$ parts are sometimes called the
{\it anomalous gravitomagnetic moment}. More precisely, $B_{20}^{u+d}(0)$
and $B_{20}^g (0)$ respectively stand for the quark and gluon contribution
to the anomalous magnetic moment of the nucleon.

Now, our first important observation is that total nucleon AGM
vanishes identically :
\begin{equation}
 B_{20}^{u+d} (0) \ + \ 
 B_{20}^g (0) \ = \ 0 .
\end{equation}
This is an exact field theoretical identity,
since it just follows from the familiar total momentum and spin sum
rules of the nucleon :
\begin{eqnarray}
 A_{20}^{u+d} (0) + A_{20}^g (0) \ \ = \ 
 {\langle x \rangle}^{u+d} \ + \ 
 {\langle x \rangle}^g &=& 1 \ \ : \ \ 
 \mbox{momentum sum rule} ,\\
 A_{20}^{u+d} (0) + B_{20}^{u+d} (0) + 
 A_{20}^g (0) + B_{20}^g (0) &=& 1 \ \ : \ \
 \mbox{spin sum rule} .
\end{eqnarray}
To proceed further, we must distinguish 3 possibilities.
\begin{eqnarray*}
 (1) &\,& B_{20}^{u+d}(0) \ = \ - \,B_{20}^g(0)
 \ \neq \ 0 ,\\
 (2) &\,& B_{20}^{u+d}(0) \ = \ B_{20}^g(0) \ = \ 0 ,\\
 (3) &\,& B_{20}^{u}(0) \ = \ B_{20}^d (0) \ = \ 
 B_{20}^g (0) \ = \ 0 .
\end{eqnarray*}

It is interesting to see that the recent lattice simulation by the
LHPC Collaboration support the 2nd possibility, i.e. the absence of
the net quark contribution to the nucleon AGM \cite{LHPC03} :
\begin{equation}
 B_{20}^{u + d}(0) 
 \ = \ 0 , \ \ \ 
 (\,\mbox{and} \ \ B_{20}^g (0) \ = \ 0 \,) , \label{NoAGM}
\end{equation}
while it at least denies the 3rd possibility, which was indicated
by Teryaev on the basis of the equivalence principle some years
ago \cite{Teryaev99}.
In fact, the result of the LHPC Collaboration for
the difference of the $u$- and $d$-quark contributions
to the nucleon anomalous gravitomagnetic moment
shows that it is clearly nonzero and has a sizable magnitude.
However, they also find that sum of the $u$- and $d$-quark
contributions, i.e. the net quark contribution to the nucleon
AGM is consistent with zero within the numerical errors.

In the following argument, we accept the relation (\ref{NoAGM}) as
a {\it theoretical postulate}. Once accepting it, Ji's
sum rule reduces to an extremely simple relation as follows,
\begin{equation}
 2 \,J^{u+d} \ = \ 
 {\langle x \rangle}^{u+d} ,
\end{equation}
which dictates the equal partition of the momentum and total
angular momentum of quark fields in the nucleon as advocated by
Teryaev \cite{Teryaev99}.

Now we can reach more surprising conclusion, based only upon two
already known empirical information at low energies \cite{W05}.
The 1st observation is that the quark and gluon fields shares
about $70 \%$ and $30 \%$ of the total nucleon momentum at low
energy scale of nonperturbative QCD :
\begin{equation}
 {\langle x \rangle}^{u+d} \ \simeq \ 0.7 ,
 \ \ \ \ \ 
 {\langle x \rangle}^g \ \simeq \ 0.3 .
\end{equation}
This can, for example, be convinced from the famous
GRV fit of the unpolarized PDF at the NLO \cite{GRV98}.
Given below is
their gluon density given at the low energy scale around
$600 \,\mbox{MeV}$ :
\begin{equation}
 x g(x, \mu_{NLO}^2) = 20.8 x^{1.6} \,(1-x)^{4.1}
 \ \ \ \mbox{at} \ \ \ 
 Q_{ini}^2 = \mu_{NLO}^2 \simeq
 (630 \,\mbox{MeV})^2 .
\end{equation}
Using it, one finds that the momentum fraction
carried by gluons is just about $30 \%$ at this energy scale : 
\begin{equation}
 {\langle x \rangle}^g \ = \ \int_0^1 x \,
 g(x,\mu_{NLO}^2) \,dx \ \simeq \ 0.3 \ \ \ \
 (30 \%) .
\end{equation}
This conversely means that, at this low energy, the quark fields
carry about $70 \%$ of total nucleon momentum and also the total
angular momentum : 
\begin{equation}
 2 \,J^{u+d} \ = \ {\langle x \rangle}^{u+d}
 \ \simeq \ 0.7 .
\end{equation}

The 2nd observation is nothing but the celebrated EMC observation
combined with the results of the subsequent polarized DIS experiments,
which revealed that the quark spin fraction is
only from $20 \%$ to $35 \%$ : 
\begin{equation}
 \Delta \Sigma \ \simeq \ 
 (0.2 \sim 0.35 ) 
 \ \ \ : \ \ 
 \mbox{weakly scale-dependent} .
\end{equation}
Combining these two observations, we are
then inevitably led to the conclusion that {\it the quark orbital angular
momentum carries nearly half of the nucleon spin at the low
energy scale around} $Q^2 \simeq (600 \,\mbox{MeV})^2$ !
\begin{equation}
 2 \,L^{u+d} 
 \ = \ 2 \,J^{u+d}
 \ - \ \Delta \Sigma
 \ \ \simeq \ \ 
 (0.35 \sim 0.5) .
\end{equation}

 \section{\large{Unpolarized GPDs and quark angular momentum
 distributions}}

Next, we turn to the discussion of the unpolarized GPD, which contains
more rich information than the corresponding generalized form factors.
Given below is the standard definition of the unpolarized GPDs
$H (x,\xi,t)$ and $E(x,\xi,t)$ :
\begin{eqnarray}
 &\,& \int \frac{d \lambda}{2 \pi} \,
 e^{i \lambda x} \,
 \langle P',s' | \bar{\psi} 
 \left( - \frac{\lambda n}{2} \right) \,\not\!n \,
 \psi \left( \frac{\lambda n}{2} \right) |
 P, s \rangle \\
 &=& \bar{U} (P',s') \,\left[ \,
 H (x,\xi,t)
 \not\!n + 
 E (x,\xi,t) \,
 \frac{i \sigma^{\mu \nu} n_\nu \Delta_\nu}{2 M} \,
 \right] \,U(P,s) .
\end{eqnarray}
As is widely known, the spin decomposition of the above amplitude is
most conveniently carried out in the Breit frame :
\begin{eqnarray}
 H_E (x,\xi,t) &\equiv& H(x,\xi,t) \ + \ 
 \frac{t}{4 \,M_N^2} \, E(x,\xi,t) , \\
 E_M (x,\xi,t) &\equiv& H(x,\xi,t) \ + \ 
 E(x,\xi,t) .
\end{eqnarray}
It corresponds to the Sachs decomposition of
the nucleon electromagnetic form factors. In fact, the 1st moment
of $H_E$ gives the Sachs electric form factor, while the 1st moment
of $E_M$ does the Sachs magnetic form factor :
\begin{eqnarray}
 \int_{-1}^1 \,H_E (x,\xi,t) \,dx &=& G_E (t) \ \ \,:
 \ \ \,
 \mbox{electric F.F.} , \\
 \int_{-1}^1 \ E_M (x,\xi,t) \,dx &=& G_M (t) \ \ :
 \ \ 
 \mbox{magnetic F.F.} .
\end{eqnarray}

Very recently, the forward limit of $E_M (x,\xi,t)$ was predicted
within the framework of the chiral quark soliton model (CQSM).
The isoscalar part was investigated by
Ossmann et. al. \cite{OPSUG05}, while the isovector part was studied
by us \cite{WT05}.
We first look into the isoscalar part, which is directly related
to the total quark contribution to the nucleon spin.
One can verify that the model satisfies the following
1st and 2nd moment sum rules :
\begin{eqnarray}
 \int \,E_M^{u+d} (x,0,0) \,dx &=& 
 3 \,(\mu_p \ + \ \mu_n ) , \\
 \int \,x \,E_M^{u+d} (x,0,0) \,dx &=&
 2 \,J^{u+d} \ = \ 1 .
\end{eqnarray}
That is, the 1st moment of $E_M^{u+d}$ gives the isoscalar magnetic moment
of the nucleon. This is an important relation, because it means that the
forward limit of the $E_M^{u+d}$ gives a distribution of nucleon
isoscalar magnetic moment in Feynman momentum $x$-space (not in ordinary
coordinate space). On the other hand, we can prove that the 2nd moment
of $E_M^{u+d}$ is reduced to twice the total quark angular momentum,
which turns out just unity in the model. This is only natutal, because
the CQSM is an effective quark model containing quark fields alone.

Fig.1(a) shows the CQSM prediction for $E_M^{u+d} (x,0,0)$.
Here, the distribution in the negative $x$ region should be
interpreted as that of antiquarks as
$E_M^{u+d}(-x,0,0) = - \,E_M^{\bar{u}+\bar{d}}(x,0,0)$ with
$x > 0$.
What is remarkable here is the $1 / x$ behavior of the
contribution of Dirac sea quarks, first pointed out by
Ossmann et.al. \cite{OPSUG05}.
It is interesting to see that, because of the peculiar antisymmetric
behavior with respect to $x$, the Dirac sea part gives no contribution
to the 1st moment sum rule, while it gives a significant contribution
to the 2nd moment, i.e. the nucleon spin sum rule.

\begin{figure}[htb] \centering
\begin{center}
 \includegraphics[width=16.0cm,height=8.0cm]{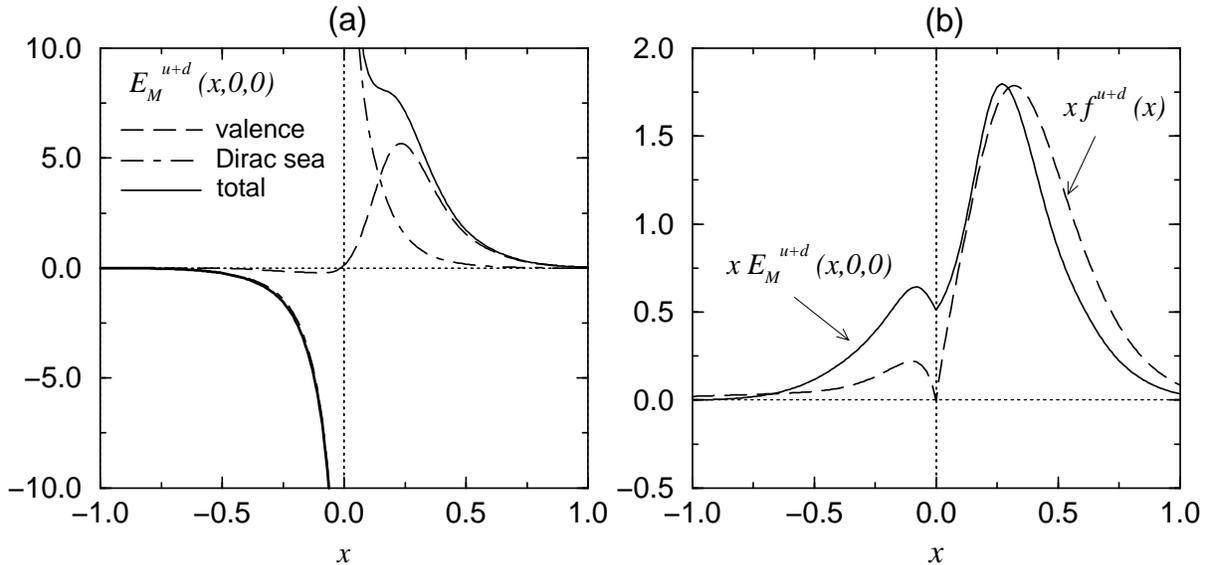}
\end{center}
\vspace*{-0.5cm}
\renewcommand{\baselinestretch}{1.20}
\caption{The theoretical prediction for $E_M^{u+d} (x,0,0)$
(a) and a comparison of the isoscalar quark spin and momentum
distributions, $x \,E_M^{u+d} (x,0,0)$ and $x \,f^{u+d}(x)$ (b).}
\label{fig:eisosca}
\end{figure}%

Here we look into the relation between the quark angular
momentum distribution and the momentum distribution in more detail.
As mentioned before, the distribution $E_M$ consists of two
parts, i.e. the familiar unpolarized distribution $f(x)$ and the genuine
or anomalous part as,
\begin{equation}
 E_M^{u+d} (x,0,0) \ \equiv \ f^{u+d}(x) + E^{u+d} (x,0,0) .
\end{equation}
Using Ji's unintegrated sum rule, the quark spin and momentum
distributions, i.e. $J^{u+d}(x)$ and $x \,f^{u+d}(x)$, are then
related as,
\begin{equation}
 2 \,J^{u+d} (x) \ = \ x \,f^{u+d}(x) + x \, E^{u+d} (x,0,0) .
\end{equation}
That is, the anomalous part gives the measure of the difference
of these two distributions. Here, we recall the important constraints
for the anomalous part of distribution.
Its first moment is proportional to the isoscalar
anomalous magnetic moment of the nucleon, which is empirically
known to be quite small : 
\begin{equation}
 \int \,E^{u+d} (x,0,0) \,dx \ = \  
 3 \,(\kappa^p + \kappa^n) \ \ : \ \ \mbox{small} .
\end{equation}
On the other hand, its 2nd moment gives
the isoscalar AGM, which vanishes exactly within the CQSM : 
\begin{eqnarray}
 \int \,x \,E^{u+d} (x,0,0) \,dx \ = \  0 \ \ : \ \ 
 \mbox{absence of AGM} .
\end{eqnarray}
Very interestingly, one observes that, while the $x$-distribution of
isoscalar anomalous magnetic is nonzero (though
small), it gives no net contribution to the total nucleon spin.
In other words, {\it the net quark contribution to the nucleon spin
is solely determined by the familiar unpolarized quark
distribution} $f^{u+d} (x)$, which can also be interpreted as
the canonical part of the isoscalar magnetic moment distribution
in the Feynman $x$-space. We emphasize that this conclusion is
never restricted to the CQSM. It would be intact also in real QCD,
since it is equivalent to assuming Eq.(\ref{NoAGM}), i.e. the absence
of the net quark contribution to the nucleon AGM.
One can also convince from Fig.1(b) that,
because of the smallness of anomalous part of distribution
$E^{u+d}(x,0,0)$, the difference of the quark spin and
momentum distributions is not very large.

Now we turn to the discussion of the isovector part. The model
expression for the isovector distribution
satisfies the desired 1st moment sum rule, that is, it reproduces the
known theoretical expression for the nucleon isovector magnetic
moment :
\begin{equation}
 \int_{-1}^1 \,E_M^{u-d} (x,0,0) \,dx \ = \ 
 - \,\frac{M_N}{9} \,N_c \,
 \sum_{n \in occ} \,\langle n | \,
 (\mbox{\boldmath $x$} \times 
 \mbox{\boldmath $\alpha$}) \cdot 
 \mbox{\boldmath $\tau$} \,| n \rangle
 \ = \ \mu_p - \mu_n .
\end{equation}
%

\begin{figure}[htb] \centering
\begin{center}
 \includegraphics[width=16.0cm,height=8.0cm]{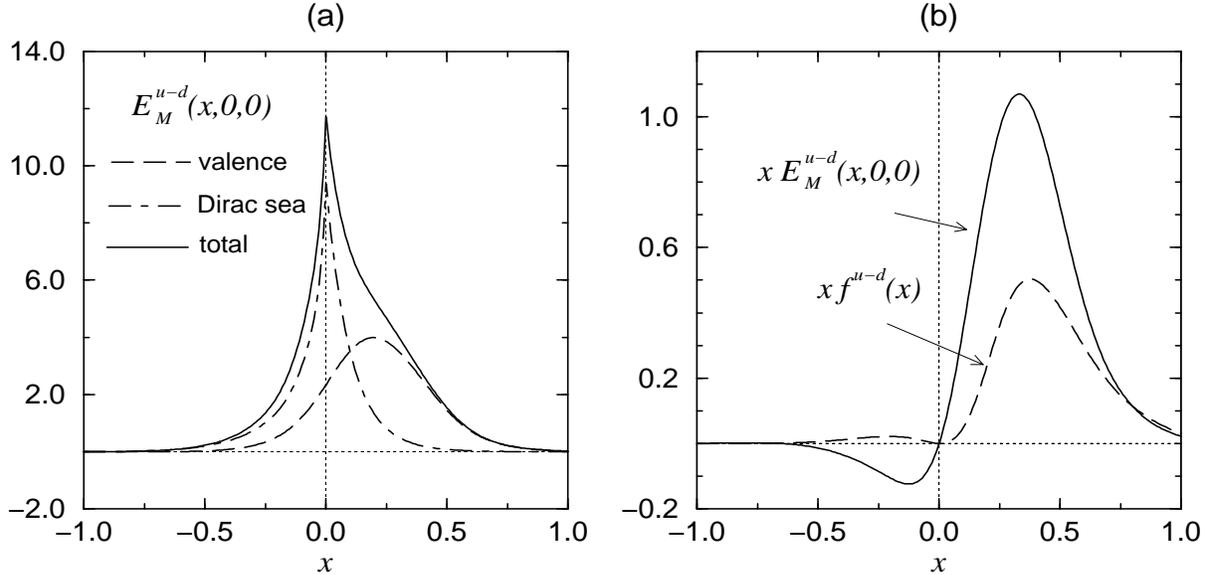}
\end{center}
\vspace*{-0.5cm}
\renewcommand{\baselinestretch}{1.20}
\caption{The theoretical prediction for $E_M^{u-d} (x,0,0)$ (a) and
a comparison of the isovector quark spin and momentum distributions,
$x \,E_M^{u-d} (x,0,0)$ and $x \,f^{u-d}(x)$ (b).}
\label{fig:eisovec}
\end{figure}%

Fig.2(a) shows the CQSM prediction for the distribution
$E_M^{u-d}(x,0,0)$. The long-dashed curve is the contribution of
three valence quarks, while the dash-dotted curve stands for the
contribution of the polarized Dirac-sea quarks, while their sum
is represented by the solid curve.
Here, a prominent feature of the CQSM predictions for the isovector
distribution is that the contribution of polarized Dirac-sea
quarks has a large and sharp peak around $x = 0$. 
What does it mean ? Since the partons with $x$ being 0 are at rest
in the longitudinal direction, its large contribution to the
magnetic moment must come from the motion of quarks and antiquarks
in the transverse plane. If one remembers the important role of the
pion clouds in the isovector magnetic moment of the nucleon, the
above transverse motion can be interpreted as simulating the
{\it pionic quark-antiquark excitation} with long-range tail.
The validity of the proposed physical picture would be verified more
clearly if one can experimentally determine the so-called impact
parameter dependent parton distribution proposed by Burkardt and
others \cite{B00}.

Next, we compare the spin and momentum distribution
in the isovector case. Assuming Ji's relation also in this case,
the measure of the difference between the spin and momentum
distribution is again given by the genuine or anomalous part of
the distribution $E^{u-d}(x,0,0)$ as
\begin{equation}
 2 \,J^{u-d} (x) \ = \ x \,f^{u-d}(x) + 
 x \, E^{u-d} (x,0,0) .
\end{equation}
Here, we find a big difference with the isoscalar case. As is clear
from the 1st moment sum rules or the magnetic moment sum rules,
\begin{eqnarray}
 \int \,f^{u-d} (x) \,dx &=& 1 \hspace{15mm} : \ \ \mbox{small} , \\
 \int \,E^{u-d} (x,0,0) \,dx &=& 
 \kappa^p - \kappa^n \ \ \ \ \ : \ \ \mbox{large} ,
\end{eqnarray}
the magnitude of the anomalous part is much larger than the canonical
charge part here.  Accordingly,
one would expect that the difference of the spin and momentum
distribution is fairly large in the isovector case.
As shown in Fig.2(b), our theoretical calculation confirms
that this is indeed the case.

 \section{\large{Summary}}

$L_q$ or $\Delta g$ ? There has been long-lasting dispute over
this issue. Here we advocated a viewpoint
which favors the importance of $L_q$. In fact, relying only upon the
following information, i.e. Ji's quark angular momentum sum rule,
the probable absence of the flavor singlet quark AGM, and the
empirical PDF information evolved down to the low energy scale,
we are inevitably led to the conclusion that the quark orbital
angular momentum carries nearly half of the total nucleon spin
at the low energy scale of nonperturbative QCD.
Note that this is a model-independent conclusion, although
the result is consistent with the prediction of the CQSM.

Naturally, for more definite confirmation, experimental extraction of
the unpolarized spin-flip GPD, at least its forward limit, is
indispensable. I stress that these forward distributions are
interesting themselves, because they give the distribution of the
nucleon anomalous magnetic moments in Feynman momentum space.
Also desirable is experimental extraction of impact-parameter
dependent parton distributions, which would certainly contain more
detailed information not only on the origin of nucleon spin but
also on the origin of the anomalous magnetic moments of a relativistic
composite particle. Can we really see chiral enhancement near
$x=0$ or large $b_\perp$?





 \bigskip

 {\small The present talk is partially based on the collaboration with
H.~Tsujimoto. The work is supported in part by a Grant-in-Aid for
Scientific Research for Ministry of Education, Culture, Sports,
Science and Technology, Japan (No. C-16540253).
}

 \end{document}